\documentstyle[psfig]{mn}

\newcommand{\gtrsim}{\mathrel{\hbox{\rlap{\lower.55ex \hbox {$\sim$}}
                   \kern-.3em \raise.4ex \hbox{$>$}}}}
\newcommand{\lesssim}{\mathrel{\hbox{\rlap{\lower.55ex \hbox {$\sim$}}
                   \kern-.3em \raise.4ex \hbox{$<$}}}}

\newif\ifAMStwofonts

\ifoldfss
  \ifCUPmtlplainloaded \else
    \NewTextAlphabet{textbfit} {cmbxti10} {}
    \NewTextAlphabet{textbfss} {cmssbx10} {}
    \NewMathAlphabet{mathbfit} {cmbxti10} {} % for math mode
    \NewMathAlphabet{mathbfss} {cmssbx10} {} %  "   "    "
  \fi
  \ifAMStwofonts
    \ifCUPmtlplainloaded \else
      \NewSymbolFont{upmath} {eurm10}
      \NewSymbolFont{AMSa} {msam10}
      \NewMathSymbol{\upi}     {0}{upmath}{19}
      \NewMathSymbol{\umu}     {0}{upmath}{16}
      \NewMathSymbol{\upartial}{0}{upmath}{40}
      \NewMathSymbol{\leqslant}{3}{AMSa}{36}
      \NewMathSymbol{\geqslant}{3}{AMSa}{3E}

    \fi
  \fi
\fi % End of OFSS

\ifnfssone
  \newmathalphabet{\mathit}
  \addtoversion{normal}{\mathit}{cmr}{m}{it}
  \addtoversion{bold}{\mathit}{cmr}{bx}{it}
  \newmathalphabet{\mathbfit} % math mode version of \textbfit{..}
  \addtoversion{normal}{\mathbfit}{cmr}{bx}{it}
  \addtoversion{bold}{\mathbfit}{cmr}{bx}{it}
  \newmathalphabet{\mathbfss} % math mode version of \textbfss{..}
  \addtoversion{normal}{\mathbfss}{cmss}{bx}{n}
  \addtoversion{bold}{\mathbfss}{cmss}{bx}{n}
  \ifAMStwofonts
    \ifCUPmtlplainloaded \else
      \UseAMStwoboldmath
      \makeatletter
      \new@mathgroup\upmath@group
      \define@mathgroup\mv@normal\upmath@group{eur}{m}{n}
      \define@mathgroup\mv@bold\upmath@group{eur}{b}{n}
      \edef\UPM{\hexnumber\upmath@group}
      \new@mathgroup\amsa@group
      \define@mathgroup\mv@normal\amsa@group{msa}{m}{n}
      \define@mathgroup\mv@bold\amsa@group{msa}{m}{n}
      \edef\AMSa{\hexnumber\amsa@group}
      \makeatother
      \mathchardef\upi="0\UPM19
      \mathchardef\umu="0\UPM16
      \mathchardef\upartial="0\UPM40
      \mathchardef\leqslant="3\AMSa36
      \mathchardef\geqslant="3\AMSa3E
    \fi
  \fi
\fi % End of NFSS release 1
\ifnfsstwo
  \DeclareMathAlphabet{\mathbfit}{OT1}{cmr}{bx}{it}
  \SetMathAlphabet\mathbfit{bold}{OT1}{cmr}{bx}{it}
  \DeclareMathAlphabet{\mathbfss}{OT1}{cmss}{bx}{n}
  \SetMathAlphabet\mathbfss{bold}{OT1}{cmss}{bx}{n}
  \ifAMStwofonts
    \ifCUPmtlplainloaded \else
      \DeclareSymbolFont{UPM}{U}{eur}{m}{n}
      \SetSymbolFont{UPM}{bold}{U}{eur}{b}{n}
      \DeclareSymbolFont{AMSa}{U}{msa}{m}{n}
      \DeclareMathSymbol{\upi}{0}{UPM}{"19}
      \DeclareMathSymbol{\umu}{0}{UPM}{"16}
      \DeclareMathSymbol{\upartial}{0}{UPM}{"40}
      \DeclareMathSymbol{\leqslant}{3}{AMSa}{"36}
      \DeclareMathSymbol{\geqslant}{3}{AMSa}{"3E}
    \fi
  \fi
\fi % End of NFSS release 2

\ifCUPmtlplainloaded \else
  \ifAMStwofonts \else % If no AMS fonts
    \def\upi{\pi}
    \def\umu{\mu}
    \def\upartial{\partial}
  \fi
\fi
\title[Cygnus X-2]
  {X-ray timing behaviour of Cygnus X-2 at low intensities}

\author[Kuulkers, Wijnands \&\ van der Klis]
  {Erik Kuulkers$^1$\thanks{E-mail: E.Kuulkers@sron.nl (EK),
  rudy@astro.uva.nl (RW), michiel@astro.uva.nl (MvdK).},
  Rudy Wijnands$^{2\star}$ \&\ Michiel van der Klis$^{2\star}$\\
  $^1$Space Research Organization Netherlands, Sorbonnelaan 2, 3584 CA Utrecht,
  \&\ Astronomical Institute, Utrecht University, \\P.O.~Box 80000, 3507 TA
  Utrecht, The Netherlands\\
  $^2$Astronomical Institute ``Anton Pannekoek", University of Amsterdam and
  Center for High-Energy Astrophysics, \\Kruislaan 403, 1098 SJ Amsterdam,
  The Netherlands}

\date{Accepted. Received.}
\pagerange{\pageref{firstpage}--\pageref{lastpage}}
\pubyear{1999}

\begin{document}

\label{firstpage}

\maketitle

\begin{abstract}

It is known that the overall (mean) intensity of the low-mass
X-ray binary Cyg\,X-2 varies 
on time scales from a day to months, independently from the variations 
on time scales of hours to a day by which the source moves between 
the horizontal, normal and flaring branches.

We present {\it RXTE} PCA observations of 
Cyg\,X-2, taken when its overall intensity was near its lowest values,
in October 1996 and September 1997.
For the first time we perform a study of the fast timing behaviour at such
low intensities.
During the October 1996 observations the source was in the left part of the 
horizontal branch, and during the September 1997 observations most 
likely in the lower parts of the normal branch and flaring branch.

We find that the properties of the very low frequency noise during the
September 1997 observations are consistent with a monotonic decrease in 
its strength and power-law index as a function of overall intensity. 
In contrast, the strength of the 
$\sim$6\,Hz normal branch quasi-periodic oscillations 
do {\it not} vary monotonically with overall
intensity. They are strongest at medium overall  
intensity and weaker both when the overall intensity is low and when the
overall intensity is high. 

\end{abstract}

\begin{keywords}
accretion, accretion disks -- binaries: close -- stars: individual: Cygnus X-2
-- stars: neutron -- X-rays: stars
\end{keywords}

\section{Introduction}

Cygnus X-2 is one of the brightest persistent low-mass X-ray binaries. It
varies on time scales from milliseconds to months 
(e.g.\ Kuulkers, van der Klis \&\ Vaughan 1996, 
Wijnands, Kuulkers \&\ Smale 1996, Wijnands et al.\ 1997a, 1998a). 
The primary is a neutron star, while the donor star is an A9 
subgiant. They orbit each other
with a period of $\sim$9.8~days (Cowley, Crampton \&\ Hutchings 1979;
Casares, Charles \&\ Kuulkers 1998).
The mass accretion rate is high 
(\.M$\sim$10$^{18}$\,g\,s$^{-1}$), giving rise to near-Eddington
X-ray luminosities (see Smale 1998). 
The source shows the Z source behaviour in X-ray colour-colour diagrams and
hardness-intensity diagrams, and associated fast timing
($\lesssim$100\,sec) properties that are 
characteristic for such high X-ray luminosity.
It shows type~I X-ray bursts (see Smale 1998, and references therein)
and kiloHertz (kHz) quasi-periodic oscillations (QPO;
Wijnands et al.\ 1998a).
Five other persistent high luminosity neutron stars
show similar X-ray behaviour. They are referred to as 
``Z'' sources, because of the Z shape of the tracks they trace out in the colour-colour diagram
(Hasinger \&\ van der Klis 1989).

The limbs of the Z are, from top to bottom, called horizontal branch, 
normal branch and flaring branch. It is thought that 
mass-accretion rate increases from the horizontal branch, through the normal branch, to the flaring branch. 
In the horizontal branch and upper part of the normal branch QPO are 
present with frequencies varying between $\sim$15 and $\sim$60\,Hz 
(called horizontal branch QPO or HBO) together with a noise 
component below $\la$20\,Hz (called low-frequency noise or LFN). On the normal branch 
different QPO (called normal branch QPO or NBO) are present with frequencies of 5--7\,Hz. 
In the Z sources Sco\,X-1 and GX\,17+2 the normal branch QPO merge smoothly 
into flaring branch QPO (called FBO) with frequencies of up to $\sim$20\,Hz on the lower 
part of the flaring branch . No such flaring branch QPO
have been reported in Cyg\,X-2, although $\sim$26\,Hz QPO were seen when
Cyg\,X-2 was in the upper part of the flaring branch, during an intensity `dip'
(Kuulkers \&\ van der Klis 1995).

The Rossi X-ray Timing Explorer (RXTE) has opened up a new window on
low-mass X-ray binaries in the millisecond regime. KHz QPO 
(for a review see e.g.\ van der Klis 1998) 
have been detected in all 6 Z sources 
(Van der Klis et al.\ 1996, 1997, Wijnands et al.\ 1997b, 1998a, 1998b; 
Jonker et al.\ 1998; Zhang, Strohmayer \&\ Swank 1998). 
The frequency of the kHz QPO increases with increasing mass-accretion rate.

Of the Z sources, Cyg\,X-2 displays the most noticeable variations 
in the X-ray intensity on long time scales 
(days to months, e.g.\ Smale \&\ Lochner 1992; Wijnands et al.\ 1996; 
see also Kong, Charles \&\ Kuulkers 1998). These so-called 
secular variations (which have been empirically divided into three intensity 
intervals, called low, medium and high intensity state) recur on a time scale 
of $\sim$78~days
and are associated with systematic changes in 
position and shape of the Z track in the colour-colour and hardness-intensity diagrams
(Kuulkers et al.\ 1996; Wijnands et al.\ 1996; 1997a). 
They are clearly distinct from the process by which the source traces out
the Z track itself on a time scale of hours to a day.
As the source goes 
from the medium intensity to high intensity state 
(or vice versa) the fast timing properties
in at least the normal branch change (Wijnands et al.\ 1997a).

On one occasion when Cyg\,X-2 was very faint in X-rays, no clear Z pattern 
was seen in the colour-colour and hardness-intensity diagram, but only a long diagonal branch associated with 
flaring behaviour which is stronger at higher energies 
(see Kuulkers et al.\ 1996).
Since the fast timing properties of the source in the low intensity state were
unknown we proposed to observe the source in this rare state by using the
RXTE All Sky Monitor (ASM) to trigger pointed observations.
In this paper we report on the results of these observations.

\section{Observations and analysis}

The RXTE Proportional Counter Array (PCA, Bradt, Rothschild \&\ Swank
1993) obtained Target of Opportunity
observations of Cyg\,X-2 on 1996 October 31 05:29--08:00~UTC
(orbital phase according to Casares et al.\ 1998:
$\phi_{\rm orb}$$\sim$0.48--0.49, where phase zero corresponds to X-ray source 
superior conjunction),
1997 September 28 09:24--18:13~UTC ($\phi_{\rm orb}$$\sim$0.22--0.26) and
1997 September 29 04:36--13:42~UTC ($\phi_{\rm orb}$$\sim$0.30--0.34), when the
ASM rate dropped below $\sim$20\,counts\,s$^{-1}$\,SCC$^{-1}$. 
Most of the data were collected with all five proportional counter units 
(PCUs) on, simultaneously with 
a time resolution of 16\,s (129 photon energy channels, effectively covering 
2--60\,keV) and down to 16\,$\mu$s using various timing 
(event, binned and single-bit) modes covering the 2--60\,keV range. 

We constructed colour-colour and hardness-intensity diagrams from the 16\,s data using
the same energy ranges as Wijnands et al.\ (1998a). The intensity 
is defined as the 
3-PCU count rate in the energy band 2.0--16.0\,keV, 
whereas the soft and hard colours are defined as the logarithm of the 
count rate ratios between 3.5--6.4\,keV and 2.0--3.5\,keV and between 
9.7--16.0\,keV and 6.4--9.7\,keV, respectively. All count rates were
corrected for background. 

Power density spectra were made from the high time resolution data 
using 16\,s data stretches, also in the same energy range as
Wijnands et al.\ (1998a), i.e., 5.0--60\,keV. 
In order to study the low-frequency ($\la$100\,Hz) behaviour 
we fitted the 0.125--256\,Hz power spectra with a constant representing the
dead time modified Poisson noise, Lorentzians or exponentially 
cut-off power laws 
to describe peaked noise components, and a power law describing 
the underlying continuum (called very-low-frequency noise or VLFN). To search for 
kHz QPO we fitted the 256--2048\,Hz power spectra with a function described 
by a constant and a Lorentzian to describe any QPO. 
Errors quoted for the power spectral parameters were 
determined using $\Delta\chi^2$=1. Upper limits were determined using 
$\Delta\chi^2$=2.71, corresponding to 95\%\ confidence levels.

\section{Results}

\subsection{Colour-colour and hardness-intensity diagrams}

In Fig.~1 we show the colour-colour diagram and in Figs.~2 and 3 the hardness-intensity diagrams of the individual
observations (a--c) and combined (d) together with the data points of
Wijnands et al.\ (1998a). All our data points correspond to X-ray intensities
of $\sim$2000--2500\,counts\,s$^{-1}$ 
(3 PCUs), so we succeeded in catching the 
source at its lowest intensity levels.
The observations obtained in 1996 October seem to be extensions towards lower
intensity on the horizontal branch. This is apparent in both the hard hardness-intensity diagram and colour-colour diagram by comparing
with the data of Wijnands et al.\ (1998a). In 
the soft hardness-intensity diagram, however, the 1996 October data fall slightly below their horizontal branch. 

The observations obtained in 1997 September, however, 
can not be immediately placed within the general Z 
pattern behaviour of the source as defined by the Wijnands et al.\ (1998a) data.
In both hardness-intensity diagrams the 1997 September data
describe a slightly curved branch, which does not fall on top 
the earlier data points.
In the colour-colour diagram the September 29 observations trace out a curved track; the
September 28 observations fall on top of the September 29 data points and 
on top of the upper part of the normal branch of Wijnands et al.\ (1998a).
It looks as if the curved track respresents the lower part of the normal branch
and flaring branch but shifted to higher soft and hard colours. However, in the hardness-intensity diagrams no
clear indications of ``flaring'' or ``dipping'' behaviour 
(see Kuulkers et al.\ 1996) can be found.

\subsection{Power spectra}

\subsubsection{1996 October}

The mean power spectrum (5--60\,keV) of the 1996 October data 
(Fig.~4a; total of $\sim$6.3\,ksec) clearly showed 
horizontal branch QPO near $\sim$19\,Hz together with a higher harmonic near
$\sim$38\,Hz on top of a low-frequency noise component, 
confirming that the source is in the left end of the horizontal branch.
A fit to this power spectrum (LFN + 2 QPO)
resulted in a reduced $\chi^2$ of 1.72 for 114 degrees of freedom (dof).
This is not a good fit; in fact, a close inspection of the power spectrum
reveals that the harmonic is not well fitted with this model.
We therefore added another cut-off power-law component with a cut-off near
20\,Hz, i.e.\ a so-called high-frequency
noise (HFN) component, which significantly improved the fit at high 
frequencies: reduced $\chi^2=1.42$ for 112 dof. Moreover, the 
resulting fit seems to describe the second harmonic much better; the centroid 
frequency of the harmonic is 37.6$\pm$0.2, compared to 36.5$\pm$0.3 without 
the high-frequency noise component. The ratio of the harmonic frequency to the QPO frequency is 
1.96$\pm$0.01 (compare this with 1.90$\pm$0.01 without the high-frequency noise 
component). The full-width-at-half-maximum (FWHM) is $\sim$10\,Hz, compared to $\sim$20\,Hz without 
the high-frequency noise component. The resulting fit to the power spectrum including the 
high-frequency noise component is shown in Fig.~4a.

The QPO and noise
components are significant enough to divide the data up into two parts. We 
computed the S$_{\rm Z}$ values which measure the position along the Z
in the hard hardness-intensity diagram, using the Z track of Wijnands et al.\ (1998a). 
In Table 1 we give the results of fits to the power spectra 
corresponding to the
two selected regions. As Cyg\,X-2 moves further onto the horizontal branch 
(to lower inferred mass-accretion rate), the frequencies
of the horizontal branch QPO and the harmonic decrease, as expected. 

We found no evidence for kHz QPO with upper limits of $\sim$3.4\%,
when fixing the FWHM at 150\,Hz.
This is significantly different from earlier 
observations in the same part of the horizontal branch (see Section 4.3).

\subsubsection{1997 September}

The variability during both the September 28 and 29 observations was low. 
In Table 2 and Figs.~4b--d we give the results of the fits to the power 
spectra for the September 28 and 29 observations.

The mean power spectrum of the September 28 observations (total of $\sim$22.6\,ksec) 
showed weak power-law noise ($\sim$1.3\%, Fig.~4b). 
Weak QPO near $\sim$40\,Hz are, however, discernable (see inset in Fig.~4b). 
We fitted these QPO and found
that they were significant at the $\sim$4$\sigma$ level 
as estimated from an F-test for the inclusion of QPO 
[$\chi^2$/dof=104/97 vs.\ $\chi^2$/dof=130/100] and from the 68\%\ confidence 
error-scan of the integral power in the $\chi^2$-space, 
i.e.\ $\Delta\chi^2$=1. Taking into account the number of trials
decreases the significance to $\sim$3$\sigma$.
A subdivision in the colour-colour and hardness-intensity diagram tracks of this observation did not
show significant differences in the power spectral shapes.

Since the colour-colour diagram of the September 29 observations (total of $\sim$21.0\,ksec) 
indicates the presence of two different branches we decided to 
investigate the power spectra by selecting these branches as indicated by the 
two regions denoted `A' and `B' in Fig.~1b. The power spectra
are displayed in Fig.~4c and Fig.~4d, for regions `A' and `B',
respectively.
The mean power spectrum for region `A' shows only a power-law noise component
(rms $\sim$1.4\%\/),
whereas the mean power spectrum for region `B' shows a weak peaked-noise 
component between $\sim$2--20\,Hz peaking near 6--7\,Hz
(rms $\sim$3\%\/), on top of a power-law noise component (rms $\sim$1\%\/).

Since the September 28 observation and part of the September 29 observations
are parallel to the normal branch of the Wijnands et al.\ (1998a) observations, we
investigated the power spectra for normal branch QPO. None were found with upper limits
of $\sim$1.3\%\ and $\sim$0.8\%, for the September 28 and 29 observations, 
respectively, with typical values for the frequency and FWHM of 5.5\,Hz and
2.5\,Hz, respectively.
The upper limits on normal branch QPO in regions `A' and `B' of the
September 29 observation are $\sim$0.8\%\ and $\sim$1.1\%, respectively.
Upper limits on the strength of QPO in the September 29 observations,
similar to that found in the Feb 28 observations near $\sim$40\,Hz, are
$\sim$1.3\%.
We also searched the September 28 and 29 power spectra for the presence of kHz
QPO but found none. Upper limits are $\sim$3\%\ and $\sim$2.8\%, for the 
September 28 and 29 observations, respectively.

\section{Discussion}

We performed RXTE observations when the ASM indicated that 
Cyg\,X-2 was at overall low intensities.
These were sucessfully performed in October 1996 and September 1997. 
It appears that we obtained data in two different kinds of 
low intensity ``states'' on the two occasions.
In the next subsections
we discuss the two observations separately, and investigate the
kHz QPO properties.

\subsection{October 1996}

During our first observation in October 1996 we found the source in the left part 
of the horizontal branch, based on the place of the source in the colour-colour diagram and hardness-intensity diagrams and the 
presence of horizontal branch QPO at $\sim$19\,Hz and their harmonic at twice this frequency. 
Previous EXOSAT (Hasinger 1987) and 
RXTE (Focke 1996; Smale 1998) observations already showed horizontal branch QPO in the same 
frequency range. Assuming that the horizontal/normal-branch vertex is at the same 
location in the hard hardness-intensity diagram as derived by Wijnands et al.\ (1998a), 
the horizontal branch QPO frequencies found during the October 1996 observations are lower 
than expected at the same position in the Z. This most likely
indicates that the Z track of our observation was shifted with respect to 
that of Wijnands et al.\ (1998a). This is supported by the fact that our 
observation is 
located below the horizontal branch of Wijnands et al.\ (1998a) in 
the soft hardness-intensity diagram, similar to what was seen in EXOSAT data by 
Kuulkers et al.\ (1996).

We found evidence for the presence of a cut-off power-law component in the 
power 
with a cut-off frequency near 20\,Hz. A similar component has been observed
previously in Cyg X-2 (Hasinger \&\ van der Klis 1989, Wijnands et al.\ 1997a) 
and in other Z sources (Hasinger \&\ van der Klis 1989, Hertz et al.\ 1992, 
Kuulkers et al.\ 1994, 1997, Kamado, Kitamoto \&\ Miyamoto 1997), and is
mostly referred to as high-frequency noise. High-frequency noise is strongest in the horizontal branch.
We note, however, that our observed high-frequency noise strength is higher than that reported
previously for Cyg\,X-2. This may be due to the higher energy range we 
investigated
compared to that of Hasinger \&\ van der Klis (1989) and Wijnands et al.\ (1997a).
The high-frequency noise has been observed to become stronger at higher energies 
(e.g.\ Dieters \&\ van der Klis 1999).

\subsection{September 1997}

The September 1997 observations do not show clear Z behaviour in the colour-colour diagram and hardness-intensity diagrams,
although we cannot rule out the possibility that a ``complete'' 
Z was traced out on a longer timescale. However, the September 29 observations
show a curved branch in the colour-colour diagram, which might be part of the Z, 
i.e.\ the lower normal branch and 
the lower flaring branch, but shifted to higher colour values. Moreover,
the September 28 observation is aligned with the normal branch, suggesting it to be 
the same branch. It is known
that the source hardens, i.e.\ the Z-pattern shifts to higher colour values, 
when it is at overall lower intensities
(Kuulkers et al.\ 1996, Wijnands et al.\ 1996).

The situation is less clear for the hardness-intensity diagrams.
The hardness-intensity diagrams of the September observations are more reminiscent of those reported 
by Vrtilek et al.\ (1986). Such shapes are seen when the source intensity 
is at an overall low level (see Kuulkers et al.\ 1996). 
Both the hardness-intensity diagrams and colour-colour diagram of the September observations do not resemble, however,
the large diagonal branch seen with EXOSAT in 1983 (see Kuulkers et al.\ 1996),
which also occurred during a low intensity state.

For the first time we have been able to examine the rapid variability
at low overall intensities. 
We find that the very-low frequency variability during the RXTE 
September observations is low,
i.e.\ $\sim$1.0--1.4\%\ (0.1--1\,Hz, 5--60\,keV).
Such low variability was also found for the very-low-frequency noise in the medium
intensity level (1--20\,keV; Wijnands et al.\ 1997a). Our observed 
very-low-frequency-noise component is unusually flat. Its index is $\alpha$$\sim$0.6--0.7), 
consistent with extrapolating the observed
decrease in index in the normal branch (Wijnands et al.\ 1997a) from the high 
($\alpha$$\sim$1.5--1.7) to medium ($\alpha$$\sim$1) intensity level down to 
the low intensity level.

In order to compare our RXTE observations with the 1983 "diagonal branch"
observations of Cyg\,X-2 we calculated power spectra of the EXOSAT data using 
64-s data stretches. These data were obtained with a 0.25\,s time resolution 
and no energy information 
(1--20\,keV; so-called ``I3''-data from the HER3 mode, see e.g.\ Kuulkers 1995).
We used all data during which the collimator response was 100\%\
and all detectors were on source (total of $\sim$16.5\,ksec).
The resulting 0.02--2\,Hz 
average power spectrum corrected for instrumental noise,
see Berger \&\ van der Klis 1998) can be well ($\chi^2_{\rm red}$ of
1.06 for 42 dof) described by a steep power law ($\alpha$=2.0$\pm$0.2)
with 1.9$\pm$0.1\%\ rms (0.01--1\,Hz). Clearly, during the 1983 observations 
the very-low-frequency noise was much steeper than that during the September 1997 observations.

We found evidence for weak ($\sim$2\%, 5--60\,keV) QPO at $\sim$40\,Hz
during the September 28 observations. Since it has been observed in Cyg\,X-2
(Wijnands et al.\ 1997a) that the horizontal branch QPO frequency (and rms amplitude) 
{\em decreases} from the horizontal/normal-branch connection ($\sim$55\,Hz) down the normal branch 
(down to $\sim$45\,Hz), we can interpret our observed QPO 
as horizontal branch QPO occurring in the lower/middle part of the normal branch. 
The fact that the horizontal branch QPO on the normal branch has 
a similar width (i.e.\ 10--20\,Hz FWHM; e.g.\ Wijnands et al.\ 1997a, 1998a) as 
we see in our observations ($\sim$15\,Hz) supports this identification.

Since the normal branch QPO become more prominent when going from the high
to the medium intensity level, we searched for normal branch QPO in our data.
None were seen with upper limits of $\sim$1\%\ (5--60\,keV), 
which is below that seen
in the normal branch of the medium intensity level ($\sim$1--2.5\%, 1--20\,keV; 
Wijnands et al.\ 1997a).
However, when during the September 29 observations the source went from the 
inferred normal branch to the inferred flaring branch, a broad ($\sim$13\,Hz) noise component 
appeared, which
peaked near 6--7\,Hz. Interestingly, similar broad noise components have been 
reported in the lower part of the flaring branch of other observations, but with 
somewhat lower strength, i.e.\ $\sim$2\%\ (1--20\,keV; 
Hasinger \&\ van der Klis 1989;
Hasinger et al.\ 1990; Kuulkers \&\ van der Klis 1995; Wijnands et al.\ 1997a)
compared to $\sim$3\%\ (5--60\,keV).
It is apparant from Wijnands et al.\ (1997a) that the strength of these ``flaring branch QPO''
becomes stronger from the high to medium intensity level. Our observations
extend this trend to lower overall intensities.

\subsection{KiloHertz QPO}

No kHz QPO were found during the October 1996 observations with upper limits
which are significantly lower ($\sim$3\%, 5--60\,keV) than previously 
observed by
Wijnands et al.\ (1998a) in the same part of the horizontal branch as inferred from 
the horizontal branch QPO frequency (4--5\%, 5--60\,keV). It is, however, consistent with the 
upper limits quoted by Smale (1998) when the source was also in the horizontal branch
($\sim$1\%, 4--11\,keV), but at higher overall intensities. 
As noted by Smale (1998), this may indicate that the strength of kHz QPO 
(at the same position in the Z) changes as a function of the overall 
intensity level. 
Unfortunately, for our October 1996 observations we can not infer to which
overall intensity level it corresponds.

During the September 1997 observations we found no indication for kHz
QPO with upper limits of $\sim$3\%\ (5--60\,keV). 
This is consistent with the upper limits
reported previously in the normal/flaring-branch region (2--4\%, 5--60\,keV; 
Wijnands et al.\ 1998a).

\section{Conclusion}

Using RXTE we observed Cyg\,X-2 at low overall intensities, for the first
time with sufficient time resolution.
In October 1996 we found the source in the 
leftmost part of the horizontal branch. Our observations show horizontal branch QPO properties which are 
generally consistent with earlier observations in this part of the Z 
track, but also indicate significant variations in the strength of the 
kHz QPO there.
We conclude that we have seen parts of the normal branch and flaring branch during our 
September 1997 observations, when the source was seen at low overall 
intensities. They do not, however, resemble the behaviour seen during a rare 
low intensity state in 1983. Such a rare state may be observed
when the overall intensity is even lower than during our observations.
The properties of the very-low-frequency noise during our 
September low-intensity observations (low amplitude, flat power law 
slope) are consistent with extrapolation from those seen in previous 
observations at higher intensity. However, 
the lack of normal branch QPO during our observations
is {\it not} consistent with the observed trends, and suggests that 
the normal branch QPO amplitude is either non-monotonically related to intensity or 
varies independently from this parameter.

It has been suggested that obscuration by the outer accretion 
disk of the inner accretion disk regions and neutron star 
causes the low overall observed intensities during certain times
and the high to medium to low intensity level variations.
Such a configuration might be due to the precession of 
a warped accretion disk, mainly based on the rather strict periodicity
of the overall intensity variations on times scales of months
(see e.g.\ Wijnands et al.\ 1996, Wijers \&\ Pringle 1999).
We note that obscuration effectively hardens the spectrum 
which leads to the changes in the position of the Z in the colour-colour diagram
(see Kuulkers et al.\ 1996, Wijnands et al.\ 1996).
Scattering in the outer disk 
would affect the variability amplitudes by light travel time smearing 
down to a frequency of order 0.01\,Hz. While this 
picture would explain the monotonic decrease in very-low-frequency noise amplitude with 
decreasing intensity, it seems inconsistent with the flattening of its 
power law index and the non-monotonic dependence of normal branch QPO amplitude on 
intensity. A model where the low intensity states are associated with 
changes in the character of the inner accretion flow itself seems 
therefore favoured.

\section*{acknowledgements}

This work was supported in part by the Netherlands Organization for
Scientific Research (NWO) and by the
Netherlands Foundation for Research in Astronomy (ASTRON)
under grants PGS 78-277 and 781-76-017, respectively.  
EK thanks the Astronomical Institute ``Anton Pannekoek'',
where part of the analysis was done, for its hospitality.

\bsp % ``This paper has been produced using the ...''

\label{lastpage}

\newpage

{\tiny

\hspace{2cm}
\begin{table}
\begin{tabular}{c@{}p{2mm}@{}c@{}p{2mm}@{}c@{}p{2mm}@{}c@{}p{2mm}@{}c@{}p{2mm}@{}c@{}p{2mm}@{}c@{}p{2mm}@{}c@{}p{2mm}@{}c@{}p{2mm}@{}c@{}p{2mm}@{}c@{}p{2mm}@{}c@{}p{2mm}@{}c}
\multicolumn{25}{l}{\normalsize{\bf Table 1}: October 1996 power spectral fit results$^a$} \\
\multicolumn{25}{l}{~} \\
\hline
\multicolumn{1}{c}{~} & 
\multicolumn{1}{@{}c@{}}{~} & 
\multicolumn{1}{c}{~} &
\multicolumn{1}{@{}c@{}}{~} & 
\multicolumn{1}{c}{LFN} &
\multicolumn{1}{@{}c@{}}{~} & 
\multicolumn{1}{c}{~} &
\multicolumn{1}{@{}c@{}}{~} & 
\multicolumn{1}{c}{~} & 
\multicolumn{1}{@{}c@{}}{~} & 
\multicolumn{1}{c}{HBO} & 
\multicolumn{1}{@{}c@{}}{~} & 
\multicolumn{1}{c}{~} & 
\multicolumn{1}{@{}c@{}}{~} & 
\multicolumn{5}{c}{harmonic} &
\multicolumn{1}{@{}c@{}}{~} & 
\multicolumn{3}{c}{HFN} &
\multicolumn{1}{@{}c@{}}{~} & 
\multicolumn{1}{c}{~} \\
\cline{3-7}\cline{9-13}\cline{15-19}\cline{21-24}
\multicolumn{1}{c}{S$_{\rm Z}$} & 
\multicolumn{1}{@{}c@{}}{~} & 
\multicolumn{1}{c}{rms} &
\multicolumn{1}{@{}c@{}}{~} & 
\multicolumn{1}{c}{$\alpha_{\rm L}$} & 
\multicolumn{1}{@{}c@{}}{~} & 
\multicolumn{1}{c}{$\nu_{\rm L}$} &
\multicolumn{1}{@{}c@{}}{~} & 
\multicolumn{1}{c}{rms} & 
\multicolumn{1}{@{}c@{}}{~} & 
\multicolumn{1}{c}{$\Gamma$} & 
\multicolumn{1}{@{}c@{}}{~} & 
\multicolumn{1}{c}{$\nu$} & 
\multicolumn{1}{@{}c@{}}{~} & 
\multicolumn{1}{c}{rms} & 
\multicolumn{1}{@{}c@{}}{~} & 
\multicolumn{1}{c}{$\Gamma$} & 
\multicolumn{1}{@{}c@{}}{~} & 
\multicolumn{1}{c}{$\nu$} &
\multicolumn{1}{@{}c@{}}{~} & 
\multicolumn{1}{c}{rms} & 
\multicolumn{1}{@{}c@{}}{~} & 
\multicolumn{1}{c}{$\nu_{\rm H}$} & 
\multicolumn{1}{@{}c@{}}{~} & 
\multicolumn{1}{c}{$\chi^2_{\rm red}$/} \\
\multicolumn{1}{c}{~} & 
\multicolumn{1}{@{}c@{}}{~} & 
\multicolumn{1}{c}{(\%)} & 
\multicolumn{1}{@{}c@{}}{~} & 
\multicolumn{1}{c}{~} & 
\multicolumn{1}{@{}c@{}}{~} & 
\multicolumn{1}{c}{(Hz)} & 
\multicolumn{1}{@{}c@{}}{~} & 
\multicolumn{1}{c}{(\%)} & 
\multicolumn{1}{@{}c@{}}{~} & 
\multicolumn{1}{c}{(Hz)} & 
\multicolumn{1}{@{}c@{}}{~} & 
\multicolumn{1}{c}{(Hz)} & 
\multicolumn{1}{@{}c@{}}{~} & 
\multicolumn{1}{c}{(\%)} & 
\multicolumn{1}{@{}c@{}}{~} & 
\multicolumn{1}{c}{(Hz)} & 
\multicolumn{1}{@{}c@{}}{~} & 
\multicolumn{1}{c}{(Hz)} & 
\multicolumn{1}{@{}c@{}}{~} & 
\multicolumn{1}{c}{(\%)} & 
\multicolumn{1}{@{}c@{}}{~} & 
\multicolumn{1}{c}{(Hz)} & 
\multicolumn{1}{@{}c@{}}{~} & 
\multicolumn{1}{c}{dof} \\
\hline
$-$0.29$\pm$0.04 && 7.8$\pm$0.4 && $-$0.11$\pm$0.06 && 3.0$\pm$0.4 && 8.7$\pm$0.1 && 3.1$\pm$0.1 && 18.39$\pm$0.03 && 4.8$\pm$0.3 && 8.8$\pm$1.2 && 36.0$\pm$0.3 && 10.5$\pm$0.5 && 20.4$\pm$1.5 && 1.13/112 \\
$-$0.21$\pm$0.04 && 8.0$\pm$0.4 && $-$0.08$\pm$0.04 && 3.4$\pm$0.5 && 8.3$\pm$0.1 && 3.5$\pm$0.1 && 19.73$\pm$0.03 && 4.3$\pm$0.3 && 8.8$\pm$1.1 && 39.0$\pm$0.3 && 10.2$\pm$0.4 && 21.4$\pm$1.7 && 1.39/112 \\
\hline
\multicolumn{25}{l}{\footnotesize $^a$\,VLFN and LFN/HFN rms (5--60\,keV) 
integrated between 0.1--1\,Hz and 0.1--100\,Hz, respectively. $\alpha$ denotes the power-law index,} \\
\multicolumn{25}{l}{\footnotesize $^{~~}$\,$\nu$ the cut-off frequency (noise) or centroid frequency (QPO), and $\Gamma$ the FWHM (QPO).} \\
\end{tabular}
\end{table}

}

\begin{table}
\begin{tabular}{ccc@{}p{2mm}@{}cccc}
\multicolumn{8}{l}{\normalsize{\bf Table 2}: September 1997 power spectral fit results$^a$} \\
\multicolumn{8}{l}{~} \\
\hline
\multicolumn{8}{c}{~} \\
\multicolumn{8}{l}{\underline{September 28}:} \\
\multicolumn{1}{c}{~} & \multicolumn{2}{c}{VLFN} & \multicolumn{1}{@{}c@{}}{~} & 
\multicolumn{3}{c}{QPO} & 
\multicolumn{1}{c}{~} \\
\cline{2-3}\cline{5-7}
\multicolumn{1}{c}{~} & \multicolumn{1}{c}{rms} & 
\multicolumn{1}{c}{$\alpha_{\rm V}$} & \multicolumn{1}{@{}c@{}}{~} & 
\multicolumn{1}{c}{rms} & \multicolumn{1}{c}{$\Gamma$} & 
\multicolumn{1}{c}{$\nu$} & \multicolumn{1}{c}{$\chi^2_{\rm red}$/dof} \\
\multicolumn{1}{c}{~} & \multicolumn{1}{c}{(\%)} & \multicolumn{1}{c}{~} & 
\multicolumn{1}{@{}c@{}}{~} & \multicolumn{1}{c}{(\%)} & 
\multicolumn{1}{c}{(Hz)} & \multicolumn{1}{c}{(Hz)} & \multicolumn{1}{c}{~} \\
\hline
 & 1.30$\pm$0.02 & 0.61$\pm$0.04 && 1.7$\pm$0.2 & 13$^{+5}_{-2}$ & 41.0$\pm$0.7 & 1.07/97 \\
\hline
\multicolumn{8}{c}{~} \\
\multicolumn{8}{l}{\underline{September 29}:} \\
\multicolumn{1}{c}{~} & 
\multicolumn{2}{c}{VLFN} & \multicolumn{1}{@{}c@{}}{~} & 
\multicolumn{3}{c}{LFN} &
\multicolumn{1}{c}{~} \\
\cline{2-3}\cline{5-7}
\multicolumn{1}{c}{A/B} & \multicolumn{1}{c}{rms} & 
\multicolumn{1}{c}{$\alpha_{\rm V}$} &
\multicolumn{1}{@{}c@{}}{~} & 
\multicolumn{1}{c}{rms} & \multicolumn{1}{c}{$\alpha_{\rm L}$} & 
\multicolumn{1}{c}{$\nu_{\rm L}$} & \multicolumn{1}{c}{$\chi^2_{\rm red}$/dof} \\
\multicolumn{1}{c}{~} & \multicolumn{1}{c}{~} & \multicolumn{1}{c}{(\%)} &
\multicolumn{1}{c}{~} & \multicolumn{1}{c}{(\%)} & \multicolumn{1}{c}{~} &
\multicolumn{1}{c}{(Hz)} & \multicolumn{1}{c}{~} \\
\hline
A & 1.41$\pm$0.04 & 0.71$\pm$0.04 && $<$1.0 & $-$1.7$^b$ & 3.8$^b$ & 0.96/77 \\
B & 1.00$\pm$0.06 & 1.1$\pm$0.2 && 2.7$\pm$0.2 & $-$1.7$\pm$0.6 & 3.8$\pm$0.9 & 1.22/74 \\
\hline
\multicolumn{8}{l}{\footnotesize $^a$\,VLFN and LFN rms (5--60\,keV) integrated between 0.1--1\,Hz and 0.1--100\,Hz,} \\
\multicolumn{8}{l}{\footnotesize $^{~~}$\,respectively. $\alpha$ denotes the power-law index, whereas $\nu$ indicates either the} \\
\multicolumn{8}{l}{\footnotesize $^{~~}$\,cut-off frequency (noise) or centroid frequency (QPO).} \\
\multicolumn{8}{l}{\footnotesize $^b$\,Parameter fixed.} \\
\end{tabular}
\end{table}

\newpage

~\\

\newpage

\begin{figure}
\centerline{\hbox{
\psfig{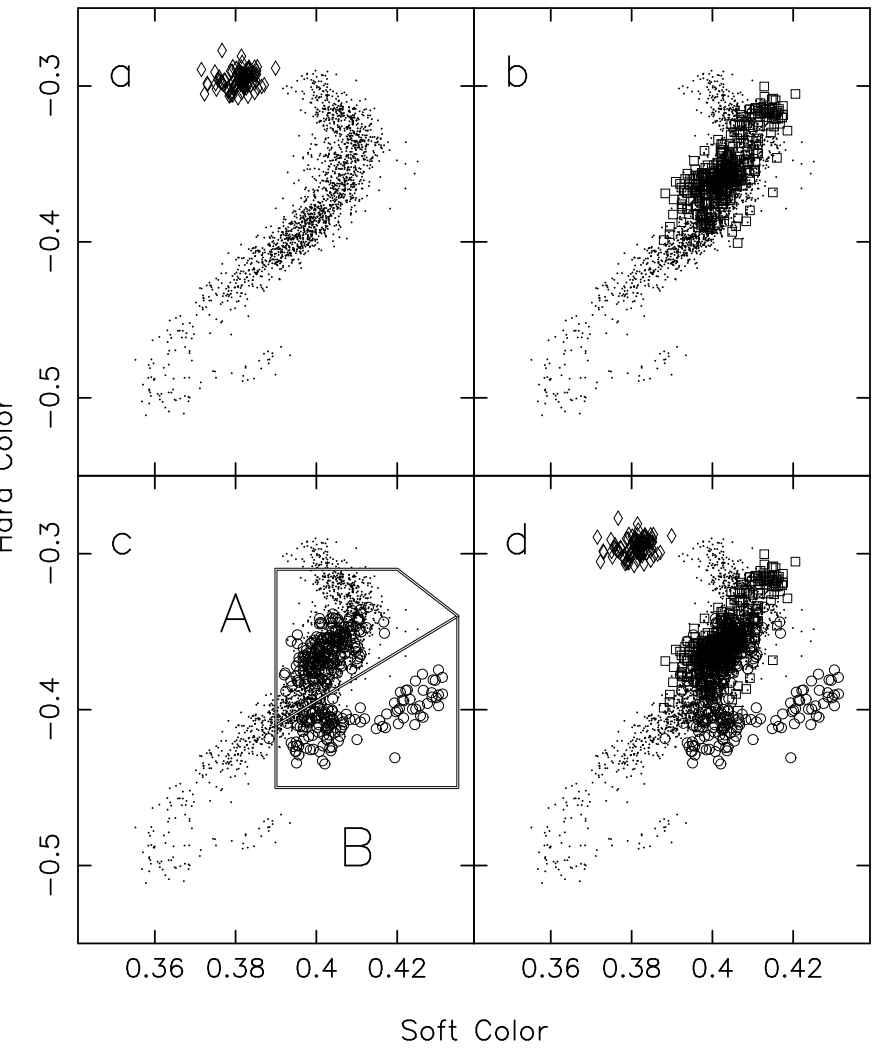}}
}
\caption{X-ray colour-colour diagram data from 1996 October 31 ({\bf a.}), 1997 
September 28 ({\bf b.}), 1997 September 29 ({\bf c.}), and all observations together 
({\bf d.}). The Wijnands et al.\ (1998a) data are represented by dots.
All points are 64\,s averages.}
%\label{heao_lc}
\end{figure}

\newpage

~\\

\newpage

~\\

\newpage

\begin{figure}
\centerline{\hbox{
\psfig{figure=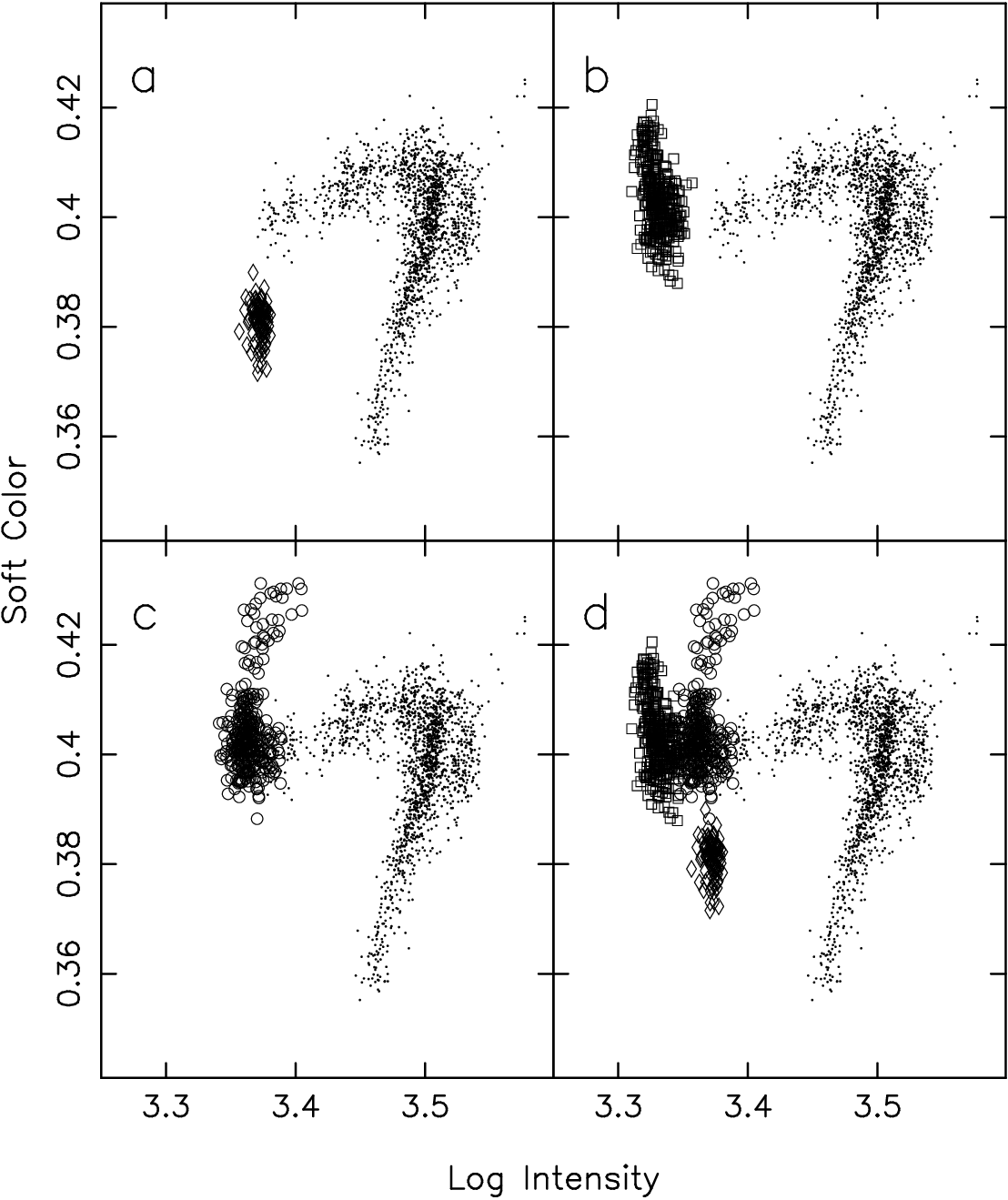,bbllx=21pt,bblly=65pt,bburx=567pt,bbury=707pt,width=12cm}}
}
\caption{Soft hardness versus intensity diagram of data from 1996 October 31 
({\bf a.}), 1997 September 28 ({\bf b.}), 1997 September 29 ({\bf c.}), and all 
observations together ({\bf d.}). The Wijnands et al.\ (1998a) data are 
represented by dots. All points are 64\,s averages.}
%\label{heao_lc}
\end{figure}

\newpage

~\\

\newpage
~\\

\newpage

\begin{figure}
\centerline{\hbox{
\psfig{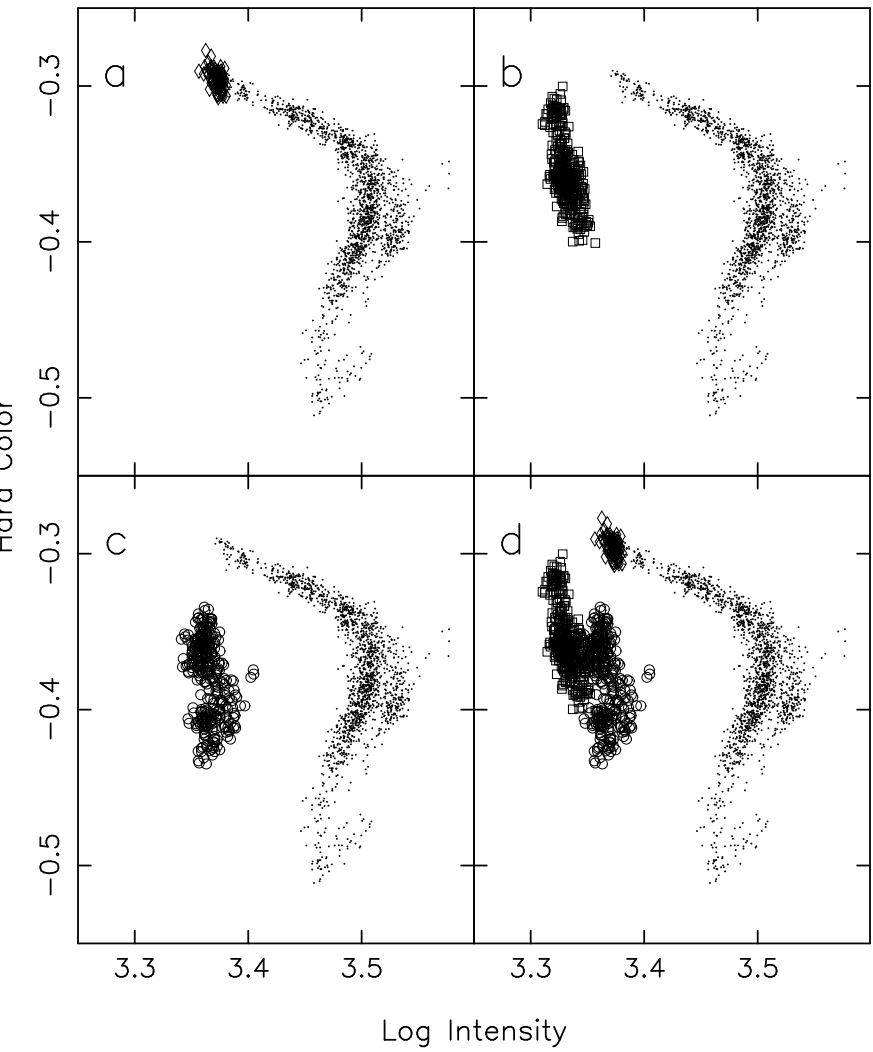}}
}
\caption{Hard hardness versus intensity diagram of data from 1996 October 31 
({\bf a.}), 1997 September 28 ({\bf b.}), 1997 September 29 ({\bf c.}), and all 
observations together ({\bf d.}). The Wijnands et al.\ (1998a) data are 
represented by dots. All points are 64\,s averages.}
%\label{heao_lc}
\end{figure}

\newpage

~\\

\newpage

~\\

\newpage

\begin{figure}
\centerline{\hbox{
\psfig{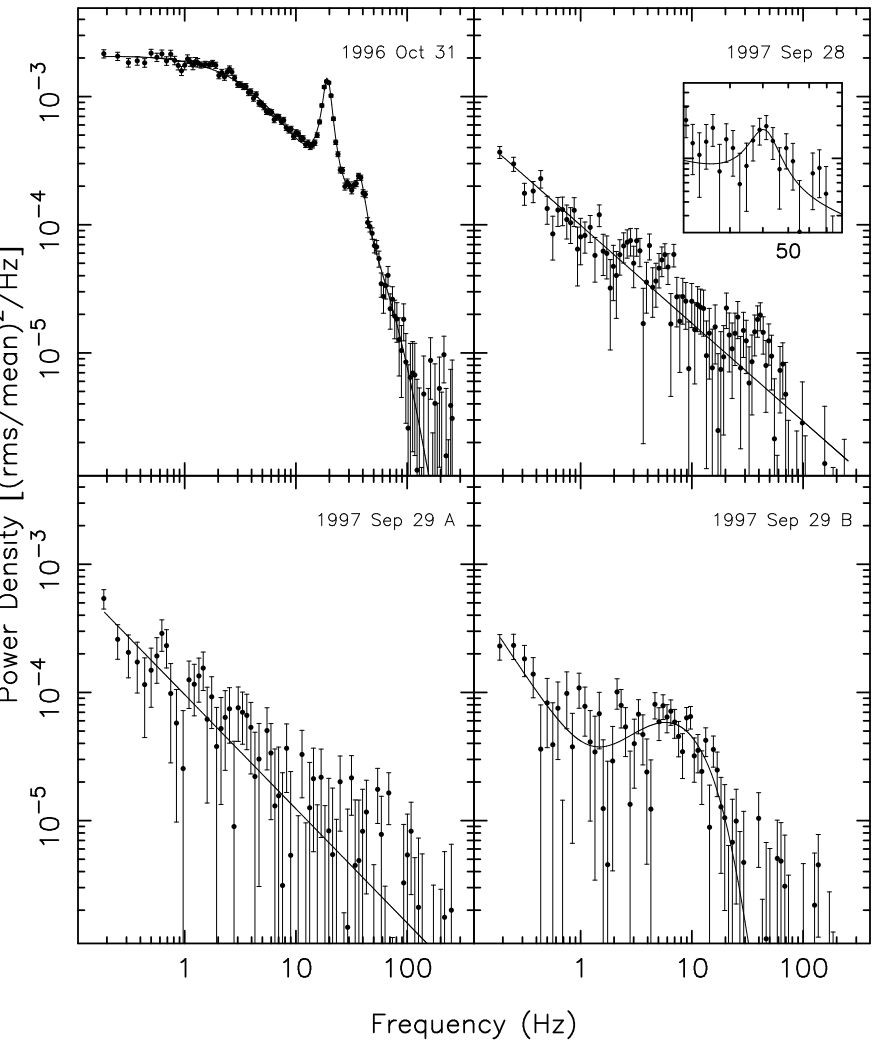}}
}
\caption{Average power spectra of data from 1996 October 31 (top left),
1997 September 28 (top right), 1997 September 29 part A, (bottom left) and
1997 September 29 part B (bottom right). The corresponding fits are shown for 
clarity.}
%\label{heao_lc}
\end{figure}

\end{document}